\documentstyle[twoside,fleqn,espcrc2,epsfig,slashed]{article}

\def\bbox{\lower0.95pt\hbox{$\Box$}}
\def\thebibliography#1{\section*{References}\list{\arabic{enumi}.}
  {\settowidth\labelwidth{#1.}\leftmargin=1.67em
   \labelsep\leftmargin \advance\labelsep-\labelwidth
   \itemsep1pt \parsep1pt
   \usecounter{enumi}}\def\makelabel##1{\rlap{##1}\hss}%
   \def\newblock{\hskip 0.11em plus 0.33em minus -0.07em}
   \sloppy \clubpenalty=4000 \widowpenalty=4000 \sfcode`\.=1000\relax}

\title{\vspace{-2.5cm}
       {\normalsize DESY 99--130}    \\[-0.2cm]
       {\normalsize HLRZ 99--39}    \\[-0.2cm]
       {\normalsize September 1999}   \\[1.25cm]
Lattice chiral fermions in the background of non-trivial
  topology\thanks{Talk presented by V. Bornyakov at Lattice '99, Pisa, Italy}} 

\author{V. Bornyakov\address{Institute for High Energy Physics IHEP,
     RU-142284 Protvino, Russia},
     A. Hoferichter\address{Deutsches Elektronen-Synchrotron DESY and NIC,
     D-15735 Zeuthen, Germany},
     G. Schierholz$^{\rm b,}$\address{Deutsches Elektronen-Synchrotron DESY,
     D-22603 Hamburg, Germany}, and
     A. Thimm$^{\rm b,}$\address{Institut f\"ur Theoretische Physik, Freie
     Universit\"at Berlin, D-14195 Berlin, Germany}}

\begin{document}

\begin{abstract} 
\noindent 
We address the problem of numerical simulations in the background
non-trivial topology in the chiral Schwinger model. An effective fermionic
action is derived which is in accord with established analytical
results, and which satisfies the anomaly equation. We describe a
numerical evaluation of baryon number violating amplitudes, specifically
the 't Hooft vertex.
\end{abstract}

\maketitle

\section{CFA to chiral gauge theories} 

The main idea of the Continuum Fermion Approach
(CFA)~\cite{AGDP,GS1,tHooft,bodwin}  
to chiral gauge theories is to discretize the gauge fields only, and to
consider the fermions in the continuum. In order to do so, we need to
extrapolate the lattice gauge fields to continuum gauge
fields~\cite{GKSW}.  

The generating functional for LH fermions in the sector of non-trivial
topology with charge $Q \leq 0 $ is defined by  
\begin{eqnarray}
Z(\eta,\bar{\eta}) &=&  
\int {\cal D}U \exp{(-S_G)}\, Z(A,\eta,\bar{\eta}), \\
Z(A,\eta,\bar{\eta}) &=&  \int {\cal D}\psi {\cal D}\bar{\psi} 
\exp{(-S_F + \bar{\psi}\eta + \bar{\eta}\psi)} \nonumber \\
 &\equiv& \exp{(-W(A))} \: \exp{(\bar{\eta} G \eta)} \nonumber \\
 & & \quad \quad \quad \quad \quad \times \prod_{i=1}^{|Q|} <\bar{\eta}
 \phi_i>,  
\end{eqnarray}
where $S_G(U)$ and $S_F(A,\bar{\psi},\psi)$ are the lattice gauge field action 
and the continuum fermion action, respectively, with 
$U=\exp{(\mbox{i}e\!\int\! A)}$. The $\phi_i$'s are the zero mode eigenfunctions,
and $G$ is the Green function orthogonal to the eigenfunctions' subspace.  
Whenever we write $A$ we mean the lattice gauge field extrapolated to the
continuum. 

The $U$'s live on a lattice with lattice spacing $a$. In practice the fermions
are put on a fine lattice with spacing $a_f$, with the understanding of 
finally
taking $a_f \rightarrow 0$. Gauge invariance breaking effects can be
made arbitrarily small by making $a_f/a$ small enough so that they do not
affect the continuum limit $a \rightarrow 0$ (in accordance
with~\cite{FNN}). The effective action $W(A)$ is defined
as~\cite{BST,buckow,kyoto}  
\begin{eqnarray}
W(A) &=& \lim_{a_f \rightarrow 0} W(U^f), \\
W(U^f) &=& -\log\, \det D(U^f) \nonumber \\
& &  + \mbox{local counterterms}, 
\end{eqnarray}
where $U^f$ is the link variable on the fine lattice, with (for one species of
LH fermions)
\begin{equation}
D(U^f) = \slashed{D}(U^f) P_L + \slashed{D}(1) P_R + W,
\label{chir_oper}
\end{equation}
$W$ being the Wilson term. 
For background fields with trivial topology it has been demonstrated, both
analytically and numerically, that $W(A)$ exists having all desired
properties. In particular,
\begin{equation}
\mbox{Re} W(A)  = \frac{1}{2} (W_V(A) + W_0),
\label{real}
\end{equation}
where $W_V$ ($W_0$) is the effective action of the vector (free) theory.
Furthermore, $W(A)$ gives the right anomaly, and $\mbox{Im}W(A)$
is gauge invariant in the anomaly free model. 
We also did not encounter any problems with large (singular) gauge
transformations~\cite{buckow}, which cause trouble in the overlap 
approach~\cite{NN1} (and presumably in Slavnov's approach~\cite{slavnov} as
well). 

Using Neuberger's operator~\cite{Neuberger} instead of the Dirac-Wilson
operator, $\mbox{Re}W$ can be computed on the original
lattice due to the absence of additive fermion mass renormalization.

\section{Evaluation of 't Hooft vertex}

The formal continuum expression for the 't Hooft vertex in the anomaly free 
CSM with four LH fermions of charge 1 and one RH fermion of charge 2 is
\begin{equation}
<\!H(x)\!> = \frac{\int {\cal D}A {\cal D}\psi {\cal
    D}\bar{\psi}\exp{(-S_G\!-\!S_F)}H(x)} {\int {\cal D}A {\cal D}\psi {\cal
    D}\bar{\psi}\exp{(-S_G\!-\!S_F)}},   
\label{hv1}
\end{equation}
\noindent
where 
\begin{equation}
{\cal D}\psi =\Big(\prod_{i=1}^4{\cal D}\psi_i^{(1)}\Big)\,{\cal D}\psi^{(2)},
\end{equation}
and similarly for ${\cal D}\bar{\psi}$, and where
\begin{equation}
H(x)= \frac{\pi^2}{e^4} \Big(\prod_{i=1}^{4}\psi_{i,L}^{(1)}(x)\Big)\,
\bar{\psi}_{R}^{(2)}(x) \slashed{\partial} \bar{\psi}_{R}^{(2)}(x). 
\end{equation}
The integral in the numerator is over the sector with topological charge 
$Q=-1$, while that in the denominator is over the sector with $Q=0$. 

On the $l\times l$ torus the gauge field takes the form
\begin{equation}
A_{\mu}(x) = \frac{2\pi}{l} t_{\mu} + \partial_{\mu} h(x) +
  \varepsilon_{\mu\nu} \partial_{\nu} \alpha(x) + C^Q_{\mu}(x), 
\end{equation}
where $t_{\mu}$ is the toron field, $\partial_{\mu} h$ represents the gauge
degrees of freedom, and $\varepsilon_{\mu\nu} \partial_{\nu} \alpha$ and
$C^Q_{\mu}$ are the proper dynamical fields of zero and non-zero topological
charge, respectively. 

After integration over the fermion fields and change of variables,
eq.~(\ref{hv1}) becomes
\begin{equation}
<\!H(x)\!> = \frac{\int {\cal D}\alpha {\cal D}t \exp{(-S^N_G\!-\!W_N)}
  H^0(x)}  
{\int {\cal D}\alpha {\cal D}t \exp{(-S_G\!-\!W_D)}},
\label{hv2}
\end{equation}
\noindent
where~\cite{KNN,wipf,kyoto,GS2}
\begin{eqnarray}
H^0(x) \!&\!=\!&\! \frac{\pi^2}{e^4} \Big(\prod_{i=1}^{4} 
\phi_i^{(1)}(x)\Big)\,(\phi^{(2)\dagger}_1(x)
\sigma_\mu  \partial_\mu \phi^{(2)\dagger}_2(x) \nonumber \\ 
& & \quad \quad \quad - \phi^{(2)\dagger}_2(x) \sigma_\mu  \partial_\mu 
\phi^{(2)\dagger}_1(x)). 
\label{hv3}
\end{eqnarray}
In eq.~(\ref{hv2})
\begin{equation}
S^N_G = S_G + \frac{2\pi^2}{e^2 l^2}, S_G = 
\frac{-1}{2e^2}\int \!\mbox{d}y \alpha(y) \bbox^2 \alpha(y),
\end{equation}
and 
\begin{eqnarray}
W_{N,D}(A) \!&\!=\!&\! 4 W_{N,D}^{(1)}(A) + W_{N,D}^{(2)}(A),  \\
W_D^{(k)}(A) \!&\!=\!&\! W^{(k)}(\alpha) + W^{(k)}(t),  \\
W_N^{(k)}(A) \!&\!=\!&\! W^{(k)}(\alpha) + W^{(k)}_\phi  
\end{eqnarray}
($k = 1,2$), with
\begin{equation}
W^{(k)}(\alpha) \!=\! -\frac{k^2}{4\pi}\int \mbox{d} y \alpha(y)\bbox 
\alpha(y), 
\end{equation}
\begin{equation}
W^{(k)}_\phi = -\log\Big(\Big(\frac{2k}{l^2}\Big)^{k/2}\!\!
\det\big(\!<\!\phi^{(k)}\!,\phi^{(k)}\!>\!\big) \Big). 
\label{zerom}
\end{equation}
The 't Hooft vertex $<\!H(x)\!>$ has been computed analytically in~\cite{KNN}
with the help of results obtained for the Schwinger model in~\cite{wipf}.

In this paper we present results obtained in a `hybrid' calculation.
For the fermionic expressions we use the analytically 
known formulae, which is equivalent to taking the limit $a_f=0$ in the CFA.
In this way we by-pass the costly calculation of the fermionic determinant
on ever larger lattices. 
The integration over the gauge fields is done numerically. 
It has been shown that the $t$-dependence factorizes out in both, 
the numerator and the denominator, and that the factors cancel each other 
if one adopts special fermionic boundary conditions~\cite{KNN}. 
We then find for the numerator
\begin{eqnarray}
N \!&\!=\!&\! \!\frac{C}{Z(\alpha)}
\int\! \prod_{n} \mbox{d}\alpha(n)\, \exp{\big(\frac{-\beta}{2} 
\sum_{n} (\bbox \alpha(n))^2\big)} \nonumber \\
& & \quad \quad \times \exp{(-W(\alpha)-8\alpha(x))},  
\end{eqnarray}
and for the denominator 
\begin{eqnarray}
D \!&\!=\!&\! \frac{1}{Z(\alpha)} 
\int\! \prod_{n} \mbox{d}\alpha(n)\, \exp{\big(\frac{-\beta}{2} 
\sum_{n} (\bbox \alpha(n))^2\big)} \nonumber \\
& & \quad \quad \times \exp{(-W(\alpha))},
\end{eqnarray}
where
\begin{equation}
Z(\alpha) = \int\! \prod_{n} \mbox{d}\alpha(n)\, 
\exp{\big(\frac{-\beta}{2} \sum_{n}(\bbox \alpha(n))^2\big)}
\end{equation}
and
\begin{equation}
C = \frac{64 \pi}{(m\, l)^4} \exp{\Big(-\frac{8\pi}{(m\, l)^2}\Big)}
 \eta^8(1).
\label{C}
\end{equation}
Here $m$ is the gauge boson mass, with $m^2=4e^2/\pi = 
4/a^2\pi\beta$, and $\eta$ is Dedekind's function.
The simulations were done for $m\, l = 3$ and $4$ on lattices
of size $L=l/a$ varying from $12$ to $48$. 
In Fig.~1 we show 
our results as a function of $1/L^2$. Also shown is the analytic 
result~\cite{KNN}. We find excellent agreement between our results 
extrapolated to $L = \infty$ and the analytic values.

\section{Conclusions}

We have once more illustrated that the CFA is a powerful tool for formulating
and analyzing chiral gauge theories on the lattice. 
In particular, we
find stable results at large values of $L$, which allow us to determine the
continuum numbers rather accurately. A problem of the overlap approach
is~\cite{KNN} that a Thirring term is generated dynamically, which has to be
tuned carefully so that its effective coupling vanishes. This is particularly
aggravating at large $L$, and so far has prevented a reliable extrapolation to
the continuum. In our approach, being intrinsically gauge invariant, this 
problem does not exist.
A totally numerical computation of the 't Hooft vertex is under way.

\begin{figure}[thb]
\vspace{-0.15cm}
\begin{centering}
\epsfig{figure=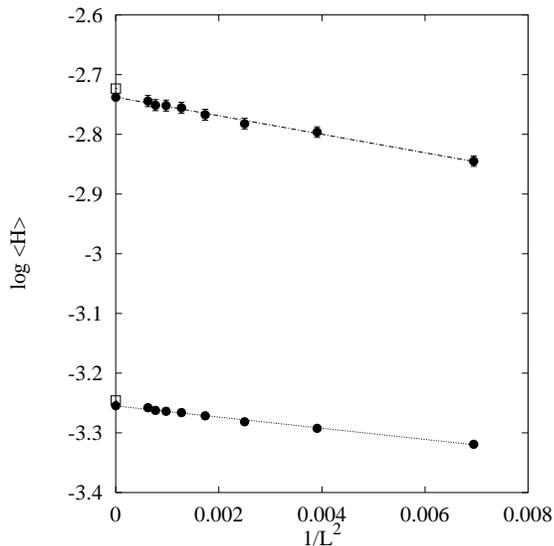,height=7.5cm,width=7.5cm}
\vspace{-0.9cm}
\caption{The 't Hooft vertex for two parameter sets, $ m l = 4$ (top) 
and 3 (bottom), together with the extrapolation to the continuum. This is 
compared to the analytic values ($\bbox$).}
\vspace{-0.7cm}
\end{centering}
\end{figure}

\end{document}